\documentclass[conference]{IEEEtran}

\usepackage{amsmath}
\usepackage{amsfonts}
\usepackage{amssymb}
\usepackage{epsf}
\usepackage{cite}
\usepackage{balance}
\usepackage{fancyhdr}
\usepackage{graphicx}
\usepackage{subfigure}
\usepackage[blocks]{authblk}
\usepackage[stable]{footmisc}
\usepackage{float}
\usepackage{fancyhdr}

\makeatletter
\def\ps@IEEEtitlepagestyle{%
  \def\@oddfoot{\mycopyrightnotice}%
  \def\@evenfoot{}%
}
\def\mycopyrightnotice{%
  {\footnotesize \copyright 2015 IEEE. Personal use of this material is permitted. Permission from IEEE must be obtained for all other uses, in any current or future media\hfill}
  \gdef\mycopyrightnotice{}
}

\newcounter{subeq}
\renewcommand{\thesubeq}{\theequation\alph{subeq}}
\newcommand{\newsubeqblock}{\setcounter{subeq}{0}\refstepcounter{equation}}
\newcommand{\mysubeq}{\refstepcounter{subeq}\tag{\thesubeq}}
\IEEEoverridecommandlockouts
\date{}
\begin{document}
\title{On the Performance of MIMO FSO Communications over Double Generalized Gamma Fading Channels}

\author[1]{Mohammadreza A.Kashani}
\author[2]{Murat Uysal}
\author[1]{Mohsen Kavehrad}

\affil[1]{Department of Electrical Engineering\authorcr
Pennsylvania State University, University Park, PA 16802\authorcr
Email: \{mza159, mkavehrad\}@psu.edu\authorcr}
\affil[2]{ Department of Electrical and Electronics Engineering \authorcr
\"{O}zye\v{g}in University,
Istanbul, Turkey, 34794 \authorcr
Email: murat.uysal@ozyegin.edu.tr\authorcr}

\maketitle
\begin{abstract}
A major performance degrading factor in free space optical communication (FSO) systems is atmospheric turbulence. Spatial diversity techniques provide a promising approach to mitigate turbulence-induced fading. In this paper, we study the error rate performance of FSO links with spatial diversity over atmospheric turbulence channels described by the Double Generalized Gamma distribution which is a new generic statistical model covering all turbulence conditions. We assume intensity modulation/direct detection with on-off keying and present the BER performance of single-input multiple-output (SIMO), multiple-input single-output (MISO) and multiple-input multiple-output (MIMO) FSO systems over this new channel model.
 \end{abstract}

\section{Introduction}\label{INTRODUCTION}
\IEEEPARstart{F}{ree-space} optical (FSO) communication enables wireless connectivity through atmosphere using laser transmitters at infrared bands. These systems provide high data rates comparable to fiber optics while they offer much more flexibility in installation and deployment. Since they operate in unregulated spectrum, no licensing fee is required leading to a cost-effective solution \cite{1,2,kashani2}.

A major performance limiting factor in FSO systems is atmospheric turbulence-induced fading (also called as scintillation) \cite{3}. Inhomogenities in the temperature and the pressure of the atmosphere result in variations of the refractive index and cause atmospheric turbulence. This manifests itself as random fluctuations in the received signal. In the literature, several statistical models have been proposed to model this random phenomenon. Historically, log-normal distribution has been the most widely used statistical model for the random irradiance experienced over atmospheric channels \cite{4,5,6,6.1,6.2,kashani3}. This model is however restricted to weak turbulence conditions and has large deviations from the experimental data when the strength of turbulence increases.

In an effort to come up with a more general model to cover a wide range of turbulence conditions, other statistical models have been proposed in the literature which include the K \cite{7}, I-K \cite{8}, log-normal Rician \cite{9}, Gamma-Gamma \cite{10}, M \cite{11} and Double Weibull \cite{12} distributions. In our recent work, we proposed the so-called Double Generalized Gamma (Double GG) as a unifying distribution for the irradiance fluctuations \cite{kashani}. This model is valid under all range of turbulence conditions and contains most of the existing models in the literature as special cases.

In our previous work \cite{kashani}, as an initial performance study, we derived the BER performance of single-input single-output (SISO) FSO link over Double GG channels. In this work, we extend our performance analysis to multiple-input multiple-output (MIMO) FSO systems. MIMO FSO systems are known to mitigate turbulence-induced fading and significantly improve the performance. Some earlier results on MIMO FSO systems over log-normal, K, negative exponential  and Gamma-Gamma channels can be found in \cite{15,20,21,Farid, Antonio}. In this paper, we study the error rate performance of single-input multiple-output (SIMO), multiple-input single-output (MISO) and MIMO FSO systems employing intensity modulation/direct detection (IM/DD) with on-off keying (OOK) over independent and not necessarily identically distributed (i.n.i.d.) Double GG turbulence channels.

The rest of the paper is organized as follows: In Section II, we introduce the MIMO FSO system model. In Section III, we provide the BER expressions for SIMO, MISO and MIMO FSO links. In Section IV, we present numerical results to confirm the accuracy of the derived expressions and demonstrate the advantages of employing spatial diversity over SISO links. Finally, Section V concludes the paper.

\section{System Model}
We consider an FSO system employing IM/DD with OOK where the information signal is transmitted via $M$ apertures and received by $N$ apertures over the Double GG channel. The received signal at the $n^{\text{th}}$ receive aperture is then given by
\begin{equation}\label{eq12}
{{r}_{n}}=\eta x\sum\limits_{m=1}^{M}{{{I}_{mn}}}+{{\upsilon }_{n}},\,\,\,n=1,\ldots ,N
\end{equation}
where $x$ represents the information bits and can be either 0 or 1, ${\upsilon }_{n}$ is the Additive White Gaussian noise (AWGN) term with zero mean and variance $\sigma _{\upsilon}^{2}={{N}_{0}}/2$ , and $\eta$ is the optical-to-electrical conversion coefficient. Here, $I_{mn}$ is the normalized irradiance from the $m^{\text{th}}$ transmitter to the $n^{\text{th}}$ receiver whose pdf follows [\citen{23}, Eq. (1)]
\begin{align}\label{eq4}
&{{f}_{I}}\left( I \right)=\frac{{{\gamma }_{2}}p{{p}^{{{\beta}_{2}}-1/2}}{{q}^{{{\beta}_{1}}-1/2}}{{\left( 2\pi  \right)}^{{1-\left( p+q \right)}/{2}\;}}{{I}^{-1}}}{\Gamma \left( {{\beta}_{1}} \right)\Gamma \left( {{\beta}_{2}} \right)}\\\nonumber
&\times G_{p+q,0}^{0,p+q}\left[ {{\left( \frac{{{\Omega }_{2}}}{{{I}^{{{\gamma }_{2}}}}} \right)}^{p}}\frac{{{p}^{p}}{{q}^{q}}\Omega _{1}^{q}}{\beta_{1}^{q}\beta_{2}^{p}}|\begin{matrix}
   \Delta \left( q:1-{{\beta}_{1}} \right),\Delta \left( p:1-{{\beta}_{2}} \right)  \\
   -  \\
\end{matrix} \right]
\end{align}
where $G_{p,q}^{m,n}\left[ . \right]$ is the Meijer’s G-function defined in [\citen{24}, Eq.(9.301)], $p$ and $q$ are positive integer numbers that satisfy ${p}/{q}\;={{{\gamma }_{1}}}/{{{\gamma }_{2}}}\;$ and $\Delta (j;x)\triangleq {x}/{j}\;,...,{\left( x+j-1 \right)}/{j}\;$, and ${{\beta }_{i}}\ge 0.5$ is a shaping parameter modeling the severity of fading. The distribution parameters ${{\gamma }_{i}}$ and ${{\Omega }_{i}}$, $i=1,2$ , of the Double GG model can be identiﬁed using the following equations
\begin{align}\label{eq9}
&{{\Omega }_{i}}={{\left( \frac{\Gamma \left( {{\beta }_{i}} \right)}{\Gamma \left( {{\beta }_{i}}+1/{{\gamma }_{i}} \right)} \right)}^{{{\gamma }_{i}}}}{{\beta }_{i}},~~i=1,2\\\label{eq8a}
\newsubeqblock
\mysubeq &\sigma _{x}^{2}=\frac{\Gamma \left( {{\beta }_{1}}+{2}/{{{\gamma }_{1}}}\; \right)\Gamma \left( {{\beta }_{1}} \right)}{{{\Gamma }^{2}}\left( {{\beta }_{1}}+{1}/{{{\gamma }_{1}}}\; \right)}-1\\\label{eq8b}
\mysubeq &\sigma _{y}^{2}=\frac{\Gamma \left( {{\beta }_{2}}+{2}/{{{\gamma }_{2}}}\; \right)\Gamma \left( {{\beta }_{2}} \right)}{{{\Gamma }^{2}}\left( {{\beta }_{2}}+{1}/{{{\gamma }_{2}}}\; \right)}-1
\end{align}
where $\sigma _{x}^{2}$ and $\sigma _{y}^{2}$ are respectively normalized variances of small and large scale irradiance ﬂuctuations.

The cumulative distribution function (cdf) of Double GG distribution can be derived from (\ref{eq4}) as \cite{kashani}
\begin{align}\label{eq5}
&{{F}_{I}}\left( I \right)=\frac{{{p}^{{{\beta}_{2}}-1/2}}{{q}^{{{\beta}_{1}}-1/2}}{{\left( 2\pi  \right)}^{{1-\left( p+q \right)}/{2}\;}}}{\Gamma \left( {{\beta}_{1}} \right)\Gamma \left( {{\beta}_{2}} \right)}\\\nonumber
&\times G_{1,p+q+1}^{p+q,1}\left[ {{\left( \frac{{{I}^{{{\gamma }_{2}}}}}{{{\Omega }_{2}}} \right)}^{p}}\frac{\beta_{1}^{q}\beta_{2}^{p}}{{{p}^{p}}{{q}^{q}}\Omega _{1}^{q}}|\begin{matrix}
   1  \\
   \Delta \left( q:{{\beta}_{1}} \right),\Delta \left( p:{{\beta}_{2}} \right),0  \\
\end{matrix} \right]
\end{align}
\section{BER Performance}
The optimum decision metric for OOK is given by \cite{21}
\begin{equation}\label{eq10}
P\left( \mathbf{r}|\text{on,}{{\text{I}}_{mn}} \right)\underset{\text{off}}{\mathop{\overset{\text{on}}{\mathop{\lessgtr }}\,}}\,P\left( \mathbf{r}|\text{off,}{{\text{I}}_{mn}} \right)
\end{equation}
where $\mathbf{r}=\left( {{r}_{1}},{{r}_{2}},...,{{r}_{N}} \right)$ is the received signal vector. Following the same approach as \cite{20,21}, the conditional bit error probabilities are given by (see \cite{20} for details of derivation)
\begin{align}\nonumber
&{{P}_{e}}(\text{off }|{{I}_{mn}})={{P}_{e}}(\text{on }|{{I}_{mn}})\\\label{eq20}
&=Q\left( \frac{1}{MN}\sqrt{\frac{{\bar{\gamma }}}{2}\sum\limits_{n=1}^{N}{{{\left( \sum\limits_{m=1}^{M}{{{I}_{mn}}} \right)}^{2}}}} \right)
\end{align}
Therefore, the average error rate can be expressed as
\begin{equation}\label{eq21}
{{P}_{\text{MIMO}}}=\int\limits_{\mathbf{I}}{{{f}_{\mathbf{I}}}}\left( \mathbf{I} \right)Q\left( \frac{1}{MN}\sqrt{\frac{{\bar{\gamma }}}{2}\sum\limits_{n=1}^{N}{{{\left( \sum\limits_{m=1}^{M}{{{I}_{mn}}} \right)}^{2}}}} \right)d\mathbf{I}
\end{equation}
where ${{f}_{\mathbf{I}}}\left( \mathbf{I} \right)$ is the joint pdf of vector $\mathbf{I}=\left( {{I}_{11}},{{I}_{12}},\ldots ,{{I}_{MN}} \right)$. The factor $M$ in (\ref{eq21}) ensures that the total transmitted powers of diversity system and SISO link are equal for a fair comparison. On the other hand, the factor $N$ is used to ensure that sum of the $N$ receive aperture areas is the same as the area of the receive aperture of the SISO link. The integral expressed in (\ref{eq21}) does not yield a closed-form solution even for simpler turbulence distributions; however, it can be calculated through numerical multi-dimensional integration. Similarly, we can use multidimensional Gaussian quadrature rule (GQR) \cite{handbook} techniques to calculate the BER for MISO case, i.e.,
\begin{equation}\label{eq22}
{{P}_{\text{MISO}}}=\int\limits_{\mathbf{I}}{{{f}_{\mathbf{I}}}}\left( \mathbf{I} \right)Q\left( \frac{\sqrt{{\bar{\gamma }}}}{M\sqrt{2}}\sum\limits_{m=1}^{M}{{{I}_{m}}} \right)d\mathbf{I}
\end{equation}
which does not yield a closed form expression either.

In the following, we focus on the SIMO case and investigate the BER performance under the assumption of optimal combining (OC) with perfect CSI where the variance of the noise in each receiver is given by $\sigma _{n}^{2}={{N}_{0}}/2N$. Therefore, replacing $M=1$ in (\ref{eq21}) we obtain
\begin{equation}\label{eq23}
{{P}_{\text{SIMO,OC}}}=\int\limits_{\mathbf{I}}{{{f}_{\mathbf{I}}}}\left( \mathbf{I} \right)Q\left( \sqrt{\frac{{\bar{\gamma }}}{2N}\sum\limits_{n=1}^{N}{I_{n}^{2}}} \right)d\mathbf{I}
\end{equation}
Eq. (\ref{eq23}) does not yield a closed-form solution and requires N-dimensional integration. Nevertheless, the Q-function can be well-approximated as $Q(x)\approx {{{e}^{-\frac{{{x}^{2}}}{2}}}}/{12}\;+{{{e}^{-\frac{2{{x}^{2}}}{3}}}}/{4}\;$ \cite{30}, and thus the average BER can be obtained as
\begin{align}\nonumber
{{P}_{\text{SIMO,OC}}}&\approx \frac{1}{12}\prod\limits_{n=1}^{N}{\int_{0}^{\infty }{{{f}_{{{I}_{n}}}}\left( {{I}_{n}} \right)}}\exp \left( \frac{-\bar{\gamma }}{4N}I_{n}^{2} \right)d{{I}_{n}}\\\label{eq24}
&+\frac{1}{4}\prod\limits_{n=1}^{N}{\int_{0}^{\infty }{{{f}_{{{I}_{n}}}}\left( {{I}_{n}} \right)}}\exp \left( \frac{-\bar{\gamma }}{3N}I_{n}^{2} \right)d{{I}_{n}}
\end{align}
The above integral can be evaluated by first expressing the exponential function in terms of the Meijer G-function presented in [\citen{29}, eq. (11)] as
\begin{equation}\label{eq25}
\exp \left( -x \right)=\operatorname{G}_{0,1}^{1,0}\left[ x\left| _{0}^{-} \right. \right]
\end{equation}
Then, a closed-form expression for (\ref{eq24}) is obtained using [\citen{29}, Eq. (21)] as
\begin{equation}\label{eq26}
{{P}_{\operatorname{SIMO},OC}}\approx \frac{1}{12}\prod\limits_{n=1}^{N}{\Lambda \left( n,4 \right)}+\frac{1}{4}\prod\limits_{n=1}^{N}{\Lambda \left( n,3 \right)}
\end{equation}
where $\Lambda \left( n,\upsilon  \right)$ is defined in (\ref{eq27}) at the top of the next page.
\begin{figure*}[t]
\begin{equation}\label{eq27}
\Lambda \left( n,\upsilon  \right)=\frac{{{\alpha }_{n}}l_{n}^{-0.5}k_{n}^{{{\beta}_{1,n}}+{{\beta}_{2,n}}}}{2{{\left( 2\pi  \right)}^{0.5\left( {{l}_{n}}-1+\left( {{k}_{n}}-1 \right)\left( {{p}_{n}}+{{q}_{n}} \right) \right)}}}G_{{{l}_{n}},{{k}_{n}}\left( {{p}_{n}}+{{q}_{n}} \right)}^{{{k}_{n}}\left( {{p}_{n}}+{{q}_{n}} \right),{{l}_{n}}}\left[ \frac{{{\left( \upsilon N \right)}^{{{l}_{n}}}}\omega _{n}^{-{{k}_{n}}}l_{n}^{{{l}_{n}}}}{{{{\bar{\gamma }}}^{{{l}_{n}}}}k_{n}^{{{k}_{n}}\left( {{p}_{n}}+{{q}_{n}} \right)}}\left| \begin{matrix}
   \Delta \left( {{l}_{n}},1 \right)  \\
   {{J}_{{{k}_{n}}}}\left( {{q}_{n}},1-{{\beta}_{1,n}} \right),{{J}_{{{k}_{n}}}}\left( {{p}_{n}},1-{{\beta}_{2,n}} \right)  \\
\end{matrix} \right. \right]
\end{equation}
\hrulefill
\end{figure*}

In (\ref{eq27}), ${{l}_{n}}$ and ${{k}_{n}}$ are positive integer numbers that satisfy ${{{p}_{n}}{{\gamma }_{2,n}}}/{2}\;={{{l}_{n}}}/{{{k}_{n}}}\;$, and ${{\operatorname{J}}_{\xi }}\left( y,x \right)$, ${{\alpha }_{n}}$ and ${{\omega }_{n}}$, $n\in \left\{ 1,2,\ldots ,N \right\}$, are defined as
\begin{align}\label{eq28c}
&{{\operatorname{J}}_{\xi }}\left( y,x \right)\\\nonumber
&=\Delta \left( \xi ,\frac{y-x}{y} \right),\Delta \left( \xi ,\frac{y-1-x}{y} \right),\ldots ,\Delta \left( \xi ,\frac{1-x}{y} \right)\\\label{eq28}
&{{\alpha }_{n}}=\frac{{{\gamma }_{2,n}}p_{n}^{{{\beta}_{2,n}}+1/2}q_{n}^{{{\beta}_{1,n}}-1/2}{{\left( 2\pi  \right)}^{{1-\left( {{p}_{n}}+{{q}_{n}} \right)}/{2}\;}}}{\Gamma \left( {{\beta}_{1,n}} \right)\Gamma \left( {{\beta}_{2,n}} \right)}\\\label{eq28b}
&{{\omega }_{n}}={{\left( {{\Omega }_{2,n}}{{p}_{n}}\beta_{2,n}^{-1} \right)}^{{{p}_{n}}}}{{\left( {{q}_{n}}{{\Omega }_{1,n}}\beta_{1,n}^{-1} \right)}^{{{q}_{n}}}}
\end{align}
The derived BER expression in (\ref{eq26}) for SIMO FSO system with OC can be seen as a generalization of BER results over other atmospheric turbulence models. For example, if we insert ${{\gamma }_{i}}=1$ and ${{\Omega }_{i}}=1$ in (\ref{eq26}), we obtain the BER expression over Gamma-Gamma channel. Setting ${{\beta }_{i}}=1$ in (\ref{eq26}), we obtain the BER for Double Weibull channel. On the other hand, for ${{\gamma }_{i}}=1$, ${{\Omega }_{i}}=1$ and ${{\beta }_{2}}=1$, (\ref{eq26}) reduces to (12) of \cite{20} reported for the K-channel. Appendix provides the details on these.

As an alternative to OC, we also consider equal gain combining (EGC) where the receiver adds the receiver branches. In this case, the average BER can be expressed as
\begin{equation}\label{eq29}
{{P}_{\text{SIMO,ECG}}}=\int\limits_{\mathbf{I}}{{{f}_{\mathbf{I}}}}\left( \mathbf{I} \right)Q\left( \frac{\sqrt{{\bar{\gamma }}}}{N\sqrt{2}}\sum\limits_{n=1}^{N}{{{I}_{n}}} \right)d\mathbf{I}
\end{equation}
It should be noted that (\ref{eq29}) is equivalent to (\ref{eq22}) obtained for the MISO FSO links. Another method is selection combining (SC) which is the least complicated of the combining schemes since it only processes one of the diversity apertures. Speciﬁcally, the SC chooses the aperture with the maximum received irradiance (or electrical SNR). Therefore, the pdf of the output of SC receiver can be obtained as
\begin{align}\nonumber
{{f}_{{{I}_{\max }}}}\left( {{I}_{\max }} \right)&=\frac{d{{F}_{{{I}_{\max }}}}\left( {{I}_{\max }} \right)}{d{{I}_{\max }}}\\\label{eq30}
&=\sum\limits_{n=1}^{N}{\prod\limits_{k=1,k\ne n}^{N}{{{f}_{{{I}_{n}}}}\left( {{I}_{\max }} \right)}}{{F}_{{{I}_{k}}}}\left( {{I}_{\max }} \right)
\end{align}
The average BER can be then expressed as
\begin{align}\nonumber
{{P}_{\text{SC}}}&=\sum\limits_{n=1}^{N}{\prod\limits_{k=1,k\ne n}^{N}{\int_{0}^{\infty }{{{f}_{{{I}_{n}}}}\left( {{I}_{\max }} \right)}}}{{F}_{{{I}_{k}}}}\left( {{I}_{\max }} \right)\\\label{eq31}
&\times \operatorname{Q}\left( {{I}_{\max }}\sqrt{\frac{{\bar{\gamma }}}{2N}} \right)d{{I}_{\max }}
\end{align}
which can be efficiently calculated through numerical means.
\begin{figure}[t]
\centering
\includegraphics[width = 8cm, height = 7.5cm]{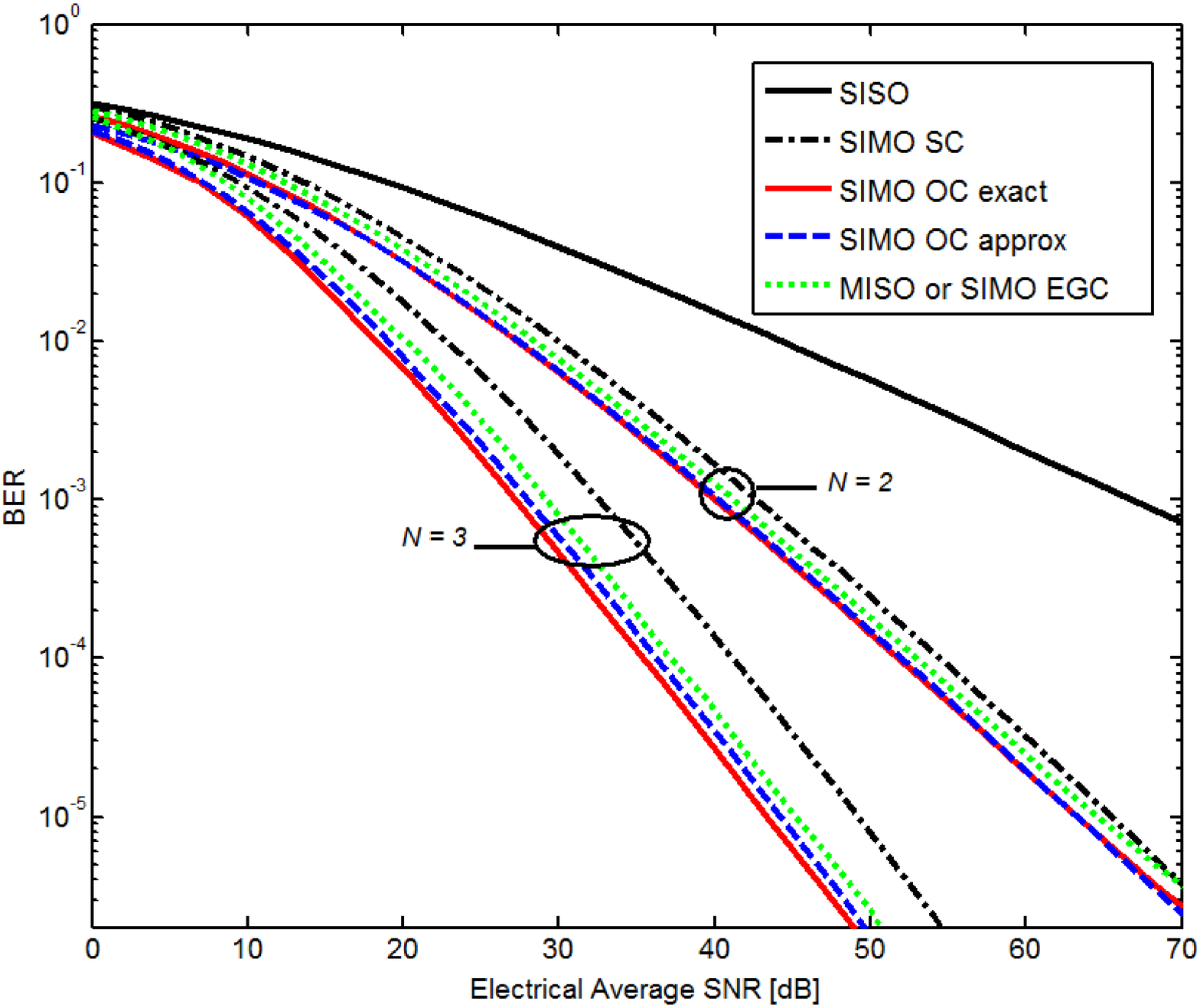}
\caption{ Comparison of the average BER between SISO and different diversity techniques for plane wave assuming i.i.d. turbulent channel defined as channel $b$.}
\end{figure}
\begin{figure}[t]
\centering
\includegraphics[width = 8cm, height = 7.5cm]{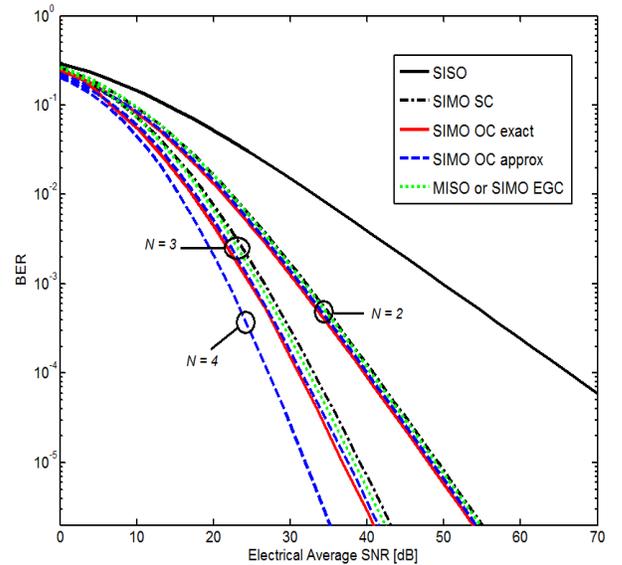}
\caption{ Comparison of the average BER between SISO and different diversity techniques for plane wave assuming i.i.d. turbulent channel defined as channel $c$.}
\end{figure}
\section{Numerical Results}
In this section, we present BER performance results of SIMO, MISO and MIMO FSO systems over Double GG channels and quantify the performance improvements over SISO systems. We consider the following four scenarios of atmospheric turbulence conditions reported in \cite{kashani}.
\begin{itemize}
   \item  \textbf{Channel a}: \emph{Plane wave and moderate irradiance fluctuations} with ${{\gamma }_{1}}=2.1690$, ${{\gamma }_{2}}=0.8530$, ${{m}_{1}}=0.55$, ${{m}_{2}}=2.35$, ${{\Omega }_{1}}=1.5793$, ${{\Omega }_{2}}=0.9671$, $p=28$ and $q=11$
  \item  \textbf{Channel b}: \emph{Plane wave and strong irradiance fluctuations} with ${{\gamma }_{1}}=1.8621$, ${{\gamma }_{2}}=0.7638$, ${{m}_{1}}=0.5$, ${{m}_{2}}=1.8$, ${{\Omega }_{1}}=1.5074$, ${{\Omega }_{2}}=0.9280$, $p=17$ and $q=7$.
  \item  \textbf{Channel c}: \emph{Spherical wave and moderate irradiance fluctuations} with ${{\gamma }_{1}}=0.9135$, ${{\gamma }_{2}}=1.4385$, ${{m}_{1}}=2.65$, ${{m}_{2}}=0.85$, ${{\Omega }_{1}}=0.9836$ and ${{\Omega }_{2}}=1.1745$, $p=7$ and $q=11$.
  \item \textbf{Channel d}: \emph{Spherical wave and strong irradiance fluctuations} with ${{\gamma }_{1}}=0.4205$, ${{\gamma }_{2}}=0.6643$, ${{m}_{1}}=3.2$, ${{m}_{2}}=2.8$, ${\Omega_{1}}=0.8336$ and ${{\Omega }_{2}}=0.9224$, $p=7$ and $q=11$.
\end{itemize}

Figs. 1-2 present the average BER over i.i.d. channels defined as channel $b$ and channel $c$ respectively. For SIMO FSO links employing OC receivers, we present approximate analytical results which have been obtained through (\ref{eq26}) along with the Monte-Carlo simulation of (\ref{eq23}). As clearly seen from Figs. 1-2, our approximate expressions provide an excellent match to simulation results. As a benchmark, the average BER of SISO FSO link is also included in these figures. It is observed that multiple receive aperture deployment signiﬁcantly improves the performance. For instance, at a target bit error rate of ${{10}^{-5}}$, we observe performance improvements of 47.2 dB and 67.2 dB for SIMO FSO links with $N=2$ and 3 receive apertures employing OC with respect to the SISO transmission over channel $b$. Similarly, for channel $c$, at a BER of ${{10}^{-5}}$, impressive performance improvements of 51.5 dB and 64.3 dB are achieved for SIMO links with $N=2$ and 3 employing OC compared to the SISO deployment. It is also illustrated that EGC receivers yield nearly the same performance as OC receivers. For example, in Fig 2, for $N=2$ the performance difference between OC and EGC receivers is merely 0.4 dB at a BER of ${{10}^{-5}}$. Also as expected, EGC and OC receivers outperform SC counterpart.

Figs. 3-4 demonstrate the BER performance of SIMO FSO links employing OC, EGC and SC receivers over non-identically distributed (i.n.i.d.) Double GG channels. Similar to i.i.d. results, our approximate closed-form expressions again yield nearly identical match to simulation results. We further compare the performance of i.n.i.d. case with respect to i.i.d. case presented in Figs1-2. For example, to achieve a BER of ${{10}^{-5}}$ in SIMO links with $N=2$ over i.n.i.d. channels $a$ and $b$,  we need 8.2 dB, 8.5 dB and 4.9 dB less in comparison to i.i.d. case respectively for OC, EGC and SC receivers. Note that in Fig. 1, we assume that both of the two channels between the transmitter and receivers are described by channel $b$. Thus, since the channel $a$ is less severe than the channel $b$, we need less SNR in comparison to i.i.d case to obtain the same BER. On the other hand, to achieve a BER of ${{10}^{-5}}$ for SIMO links with $N=2$ over i.n.i.d. channels $c$ and $d$, we need 6.8 dB more for OC and EGC receivers and 11.6 dB more for SC receiver in comparison to i.i.d channels. Note that in Fig. 2, both of the two channels between the transmitter and receivers are described by channel $c$. Therefore, as the channel $d$ is more severe than the channel $c$, we need more SNR in comparison to i.i.d case to achieve the same performance.
\begin{figure}
\centering
\includegraphics[width = 8cm, height = 7.5cm]{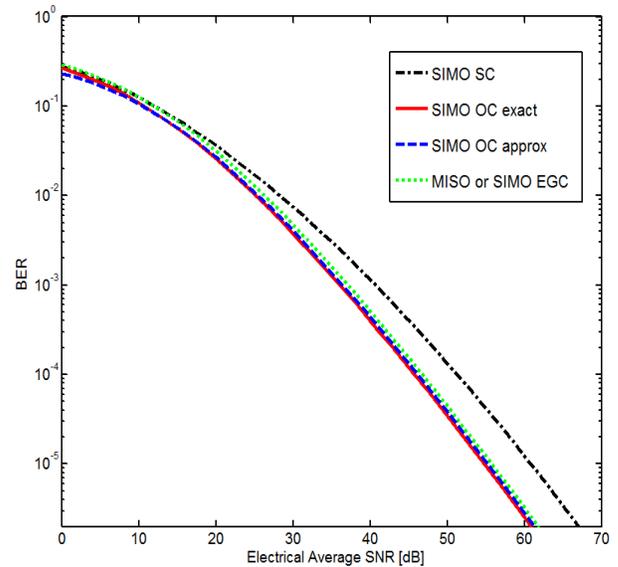}
\caption{Comparison of the OC, EGC and SC receivers for SIMO FSO links over two i.n.i.d. atmospheric turbulence channels defined as channel $a$ and channel $b$.}
\end{figure}
\begin{figure}
\centering
\includegraphics[width = 8cm, height = 7.5cm]{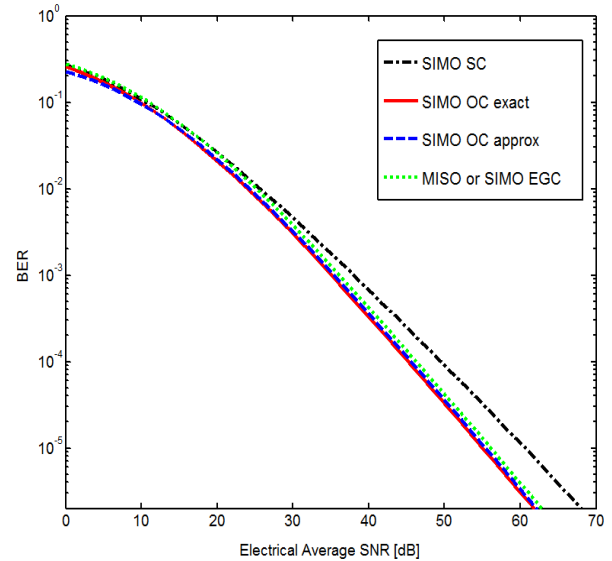}
\caption{Comparison of the OC, EGC and SC receivers for SIMO FSO links over two i.n.i.d. atmospheric turbulence channels defined as channel $c$ and channel $d$.}
\end{figure}
\section{Conclusions}\label{Con}
In this paper, we have investigated the BER performance of FSO links with spatial diversity over atmospheric turbulence channels described by the Double GG distribution. We have obtained an efficient and unified closed-form expression for the BER of SIMO FSO systems with OC receiver which generalizes existing results as special cases. For MISO and MIMO systems, we have presented BER performance based on numerical calculations of the integral expressions. Our numerical results have demonstrated that spatial diversity schemes can significantly improve the system performance and bring impressive performance gains over SISO systems. Our comparisons among SIMO FSO links employing OC, EGC and SC receivers have further demonstrated that EGC scheme presents a favorable trade-off between complexity and performance.

\section*{Appendix}\label{Appendix D}
\noindent\textbf{Proposition 1}: \emph{Inserting ${{\gamma }_{i}}=1$, ${{\Omega }_{i}}=1$ and ${{\beta }_{1}}=1$ in (\ref{eq26}), the BER of SIMO FSO links using optimal combining over K-channel is obtained.}

\noindent\textbf{Proof}: First by replacing ${{\gamma }_{i}}={{\Omega }_{i}}=p=q={{\beta }_{1}}=1$ in (\ref{eq28}) and (\ref{eq28b}), we obtain ${{\alpha }_{n}}={1}/{\Gamma \left( {{\beta }_{2,n}} \right)}$ and ${{\omega }_{n}}=\beta _{2,n}^{-1}$. Then by plugging all the values in (\ref{eq27}), we obtain
\begin{align}\nonumber
  & {{\Lambda }_{KC}}\left( n,\upsilon  \right)\\
  &=\frac{{{2}^{{{\beta}_{2,n}}-1}}}{\pi \Gamma \left( {{\beta}_{2,n}} \right)}G_{1,4}^{4,1}\left[ \frac{\left( \upsilon N \right)\beta_{2,n}^{2}}{16\bar{\gamma }}\left| \begin{matrix}
   1  \\
   \frac{1}{2},1,\frac{{{\beta}_{_{2,n}}}}{2},\frac{{{\beta}_{_{2,n}}}+1}{2}  \\
\end{matrix} \right. \right] \\\nonumber
 &=\frac{{{2}^{{{\beta}_{2,n}}-1}}}{\pi \Gamma \left( {{\beta}_{2,n}} \right)}G_{4,1}^{1,4}\left[ \frac{16\bar{\gamma }}{\upsilon N\beta_{2,n}^{2}}\left| \begin{matrix}
   \frac{1}{2},0,\frac{2-{{\beta}_{_{2,n}}}}{2},\frac{1-{{\beta}_{_{2,n}}}}{2}  \\
   0  \\
\end{matrix} \right. \right]
\end{align}
Therefore,
\begin{equation}
{{P}_{\operatorname{SIMO},OC\_KC}}\approx \frac{1}{12}\prod\limits_{n=1}^{N}{{{\Lambda }_{KC}}\left( n,4 \right)}+\frac{1}{4}\prod\limits_{n=1}^{N}{{{\Lambda }_{KC}}\left( n,3 \right)}
\end{equation}
which coincides with (21) of \cite{20}.

\noindent\textbf{Proposition 2}: \emph{Inserting ${{\gamma }_{i}}=1$, ${{\Omega }_{i}}=1$ in (\ref{eq26}), the BER of SIMO FSO links using optimal combining over Gamma-Gamma channel is obtained.}

\noindent\textbf{Proof}: First by replacing ${{\gamma }_{i}}={{\Omega }_{i}}=p=q=1$ in (\ref{eq28}) and (\ref{eq28b}), we obtain ${{\alpha }_{n}}={1}/{\Gamma \left( {{\beta }_{1,n}} \right)\Gamma \left( {{\beta }_{2,n}} \right)}$ and ${{\omega }_{n}}=\beta _{1,n}^{-1}\beta _{2,n}^{-1}$. Then by inserting all the values as well as $l=1$ and $k=2$ in (\ref{eq27}), we obtain
\begin{align}\nonumber
&{{\Lambda }_{GG}}\left( n,\upsilon  \right)=\frac{{{2}^{{{\beta }_{1,n}}+{{\beta }_{2,n}}-2}}}{\pi \Gamma \left( {{\beta }_{1,n}} \right)\Gamma \left( {{\beta }_{2,n}} \right)}\\
&\times G_{1,4}^{4,1}\left[ \frac{\upsilon N{{\left( {{\beta }_{2,n}}{{\beta }_{2,n}} \right)}^{2}}}{16\bar{\gamma }}\left| \begin{matrix}
   1  \\
   \frac{{{\beta }_{_{1,n}}}}{2},\frac{{{\beta }_{_{1,n}}}+1}{2}\frac{{{\beta }_{_{2,n}}}}{2},\frac{{{\beta }_{_{2,n}}}+1}{2}  \\
\end{matrix} \right. \right]
\end{align}
Therefore,
\begin{equation}
{{P}_{\operatorname{SIMO},OC\_GG}}\approx \frac{1}{12}\prod\limits_{n=1}^{N}{{{\Lambda }_{GG}}\left( n,4 \right)}+\frac{1}{4}\prod\limits_{n=1}^{N}{{{\Lambda }_{GG}}\left( n,3 \right)}
\end{equation}

\noindent\textbf{Proposition 3}: \emph{Inserting ${{\beta }_{i}}=1$ in (\ref{eq26}), the BER of SIMO FSO links using optimal combining over Double-Weibull channel is obtained.}

\noindent\textbf{Proof}: Inserting  ${{\beta }_{i}}=1$in (\ref{eq28}) and (\ref{eq28b}), we obtain
\begin{align}
&{{\alpha }_{n}}={{\gamma }_{2,n}}p_{n}^{3/2}q_{n}^{1/2}{{\left( 2\pi  \right)}^{1-\left( {{p}_{n}}+{{q}_{n}} \right)/2\ }}\\
&{{\omega }_{n}}={{\left( {{\Omega }_{2,n}}{{p}_{n}} \right)}^{{{p}_{n}}}}{{\left( {{q}_{n}}{{\Omega }_{1,n}} \right)}^{{{q}_{n}}}}
\end{align}

Then by plugging all the values in (\ref{eq27}), we obtain
\begin{align}\nonumber
&{{\Lambda }_{DW}}\left( n,\upsilon  \right)=\frac{{{\gamma }_{2,n}}p_{n}^{3/2}q_{n}^{1/2}l_{n}^{-0.5}k_{n}^{2}}{2{{\left( 2\pi  \right)}^{0.5\left( {{l}_{n}}-3+{{k}_{n}}\left( {{p}_{n}}+{{q}_{n}} \right) \right)}}}\\\nonumber
&\times G_{{{l}_{n}},{{k}_{n}}\left( {{p}_{n}}+{{q}_{n}} \right)}^{{{k}_{n}}\left( {{p}_{n}}+{{q}_{n}} \right),{{l}_{n}}}\left[ \frac{{{\left( \upsilon N \right)}^{{{l}_{n}}}}}{{{\left( {{\Omega }_{2,n}}{{p}_{n}}{{k}_{n}} \right)}^{{{k}_{n}}{{p}_{n}}}}}\right.\\\label{dw}
&\times \left.\frac{{{\left( {{{\bar{\gamma }}}^{-1}}{{l}_{n}} \right)}^{{{l}_{n}}}}}{{{\left( {{q}_{n}}{{\Omega }_{1,n}}{{k}_{n}} \right)}^{{{k}_{n}}{{q}_{n}}}}}\left| \begin{matrix}
   \Delta \left( {{l}_{n}},1 \right)  \\
   {{J}_{{{k}_{n}}}}\left( {{q}_{n}},0 \right),{{J}_{{{k}_{n}}}}\left( {{p}_{n}},0 \right)  \\
\end{matrix} \right. \right]
\end{align}
Thus,
\begin{equation}
{{P}_{\operatorname{SIMO},OC\_DW}}\approx \frac{1}{12}\prod\limits_{n=1}^{N}{{{\Lambda }_{DW}}\left( n,4 \right)}+\frac{1}{4}\prod\limits_{n=1}^{N}{{{\Lambda }_{DW}}\left( n,3 \right)}
\end{equation}
\balance
\bibliographystyle{IEEEtran}
\bibliography{ref}

\end{document}